\documentclass[twocolumn,tighten]{aastex62}
\shorttitle{EUV LATE PHASE IN A NON-ERUPTIVE FLARE}
\shortauthors{Dai et al.}

\begin{document}

\title{Extremely Large Extreme-ultraviolet Late Phase Powered by Intense Early Heating in a Non-eruptive Solar Flare}

\author[0000-0001-9856-2770]{Yu~Dai}
\affiliation{School of Astronomy and Space Science, Nanjing University, Nanjing 210023, People's Republic of China}
\affiliation{Key Laboratory of Modern Astronomy and Astrophysics (Nanjing University), 
Ministry of Education, Nanjing 210023, People's Republic of China}
\affiliation{Key Laboratory of Space Weather, National Center for Space Weather, China Meteorological Administration, 
Beijing 100081, People's Republic of China}

\author[0000-0002-4978-4972]{Mingde~Ding}
\affiliation{School of Astronomy and Space Science, Nanjing University, Nanjing 210023, People's Republic of China}
\affiliation{Key Laboratory of Modern Astronomy and Astrophysics (Nanjing University), 
Ministry of Education, Nanjing 210023, People's Republic of China}

\author{Weiguo~Zong}
\affiliation{Key Laboratory of Space Weather, National Center for Space Weather, China Meteorological Administration, 
Beijing 100081, People's Republic of China}

\author{Kai~E.~Yang}
\affiliation{School of Astronomy and Space Science, Nanjing University, Nanjing 210023, People's Republic of China}
\affiliation{Key Laboratory of Modern Astronomy and Astrophysics (Nanjing University), 
Ministry of Education, Nanjing 210023, People's Republic of China}

\correspondingauthor{Yu~Dai}
\email{ydai@nju.edu.cn}

\begin{abstract}
We analyzed and modeled an M1.2 non-eruptive solar flare on 2011 September 9. The flare exhibits a strong late-phase peak of the warm coronal emissions ($\sim$3~MK) of extreme-ultraviolet (EUV), with peak emission over 1.3 times that of the main flare peak. Multiple flare ribbons are observed, whose evolution indicates a two-stage energy release process. A non-linear force-free field (NLFFF) extrapolation reveals the existence of a magnetic null point, a fan-spine structure, and two flux ropes embedded in the fan dome. Magnetic reconnections involved in the flare are driven by the destabilization and rise of one of the flux ropes. In the first stage, the fast ascending flux rope drives reconnections at the null point and the surrounding quasi-separatrix layer (QSL), while in the second stage, reconnection mainly occurs between the two legs of the field lines stretched by the eventually stopped flux rope. The late-phase loops are mainly produced by the first-stage QSL reconnection, while the second-stage reconnection is responsible for the heating of main flaring loops. The first-stage reconnection is believed to be more powerful, leading to an extremely strong EUV late phase.  We find that the delayed occurrence of the late-phase peak is mainly due to the long cooling process of the long late-phase loops. Using the model enthalpy-based thermal evolution of loops (EBTEL), we model the EUV emissions from a late-phase loop. The modeling reveals a peak heating rate of 1.1~erg~cm$^{-3}$~s$^{-1}$ for the late-phase loop, which is obviously higher than previous values. 

\end{abstract}

\keywords{Sun: corona --- Sun: flares --- Sun: UV radiation --- Sun: magnetic fields --- hydrodynamics }

\section{INTRODUCTION}
Solar flares, one of the most dynamic phenomena in the solar atmosphere, are manifested as rapid and significant enhancements of  electromagnetic radiation in a wide wavelength range. It is widely accepted that solar flares are a result of rapid release of free magnetic energy stored in the solar corona, and magnetic reconnection plays a crucial role in converting the magnetic energy to plasma heating, particle acceleration, and bulk mass motions \citep{Parker63}.

According to the standard two-ribbon solar flare model, which is conventionally called the CSHKP model \citep{Carmichael64,Sturrock66,Hirayama74,Kopp76}, the evolution of a solar flare can be divided into two phases: an impulsive phase, and a following gradual phase. The impulsive phase is characterized by a rapid increase of the emissions in hard X-ray (HXR) and chromospheric lines (e.g., \ion{He}{2}), indicating a prompt response of the solar lower atmosphere to nonthermal electron bombardment and/or thermal conduction caused by the initial energy release. Since the energy transported downward cannot be effectively radiated away at the solar lower atmosphere \citep{Antiochos76}, a hot evaporative flow is consequently driven into the flare loops \citep{Antiochos78}, which then brighten up in soft X-ray (SXR) and coronal extreme-ultraviolet (EUV) lines. As the flare plasma cools down, these emissions peak sequentially in an order of decreasing temperatures, and then decay gradually toward the background levels, constituting the gradual phase \citep{Chamberlin12}.

The original CSHKP model and its variants are a two-dimensional (2D) model in nature. Nevertheless, a real solar flare takes place in a three-dimensional (3D) magnetic field, making its evolution fairly more complicated than that expected based on a simplified 2D model. By using EUV irradiance observations with the EUV Variability Experiment \citep[EVE;][]{Woods12} on board the recently launched \emph{Solar Dynamics Observatory} \citep[\emph{SDO};][]{Pesnell12} mission, \citet{Woods11} discovered a new phenomenon in some flares, namely, an ``EUV late phase". Observationally, the EUV late phase is seen as a second peak in the warm coronal emissions ($\sim$3~MK) several tens of minutes to a few hours after the \emph{GOES} SXR peak. There are, however, no significant enhancements of the SXR or hot coronal emissions ($\sim10$~MK or higher) in the EUV late phase, and spatially resolved imaging observations such as those from the Atmospheric Imaging Assembly \citep[AIA;][]{Lemen12} also on board \emph{SDO} reveal that the secondary late-phase emission comes from another set of flare loops higher and longer than the main flaring loops.

In a preliminary statistics of 25 EUV late-phase solar flares occurring during the first year of \emph{SDO} normal operations, \citet{Woods11} found that about half of them took place in two active regions (ARs), implying a specific magnetic configuration of the ARs in which EUV late-phase flares are preferentially produced. In case studies, photospheric magnetograms of the flare-hosting ARs reveal that the EUV late-phase flares are commonly involved in a multipolar magnetic field, which exhibits either a symmetric or asymmetric quadrupolar configuration \citep{Hock12,LiuK13},  or a parasitic polarity embedded in a large-scale bipolar magnetic field \citep{Dai13,SunX13,Masson17}. Such a magnetic configuration facilitates the existence of two sets of loops that are magnetically related but distinct in length, as further confirmed by the force-free coronal magnetic field extrapolations \citep{SunX13,LiYD14,Masson17}.

The EUV irradiance output from the Sun can drive disturbances in Earth's ionosphere and thermosphere, particularly during solar flares \citep{Kane71}. Hence, the origin of EUV late phase in a solar flare is of interest for its potential geo-effectiveness. Since many theoretical works on the cooling of flare plasmas have shown that the cooling time of a flare loop increases as the loop length increases \citep[e.g.,][]{Cargill95}, it was naturally proposed that the delayed  occurrence of an EUV late phase is the result of a long-lasting cooling process in the long late-phase loops initially heated simultaneously with the main flaring loops \citep{LiuK13,Masson17}. In this sense, the late phase is actually the gradual phase in long flare loops. Nevertheless, there is also growing evidence for the existence of a delayed secondary heating, which might be responsible for the production of the EUV late phase \citep{Hock12,Dai13}. In fact, the two mechanisms are not mutually exclusive, and in some EUV late-phase flares they may both play a role \citep{SunX13}. 

By using the model called enthalpy-based thermal evolution of loops \citep[EBTEL;][]{Klimchuk08,Cargill12,Barnes16}, \citet{LiYD14} and \citet{Dai18} numerically probed the production of EUV late-phase flares. In particular, \citet{Dai18} found that even with an equal energy partition between the late-phase loop and main flaring loop, the warm coronal late-phase peak is still significantly lower than the corresponding main flare peak. This result may offer a clue as to why EUV late-phase flares are rare among all solar flares. Nevertheless, as shown in the statistics in \citet{Woods11}, the ratio of the late-phase peak to main flare peak in their sample can be as high as 4. \citet{LiuK15} have recently reported a long-duration but non-eruptive flare in which the late-phase peak is $\sim2.1$ times higher than the main flare peak. For this reason they referred to it as an ``extremely large EUV late phase". By analyzing the kinematic and thermodynamic evolution of the flare, \citet{LiuK15} proposed that an ``erupted-but-failed" hot structure, which is most presumably a magnetic flux rope, serves as a persistent heating agent for this extremely large EUV late phase. \citet{WangYM16} argued that a strong constraint of the overlying arcades may prevent a flux rope from escaping, and cause the energy carried by the flux rope to be re-deposited into the thermal emissions that form a stronger late phase. This argument was further validated through a comparative study of 12 EUV late-phase flares \citep{WangYM16} showing that the relative late-phase peaks of the non-eruptive flares are systematically stronger than those of the eruptive flares.

In addition to the persistent heating scenario proposed by \citet{LiuK15}, we are wondering if an impulsive heating on the late-phase loops, which is temporally close to the main flare heating, can also produce a significantly large EUV late phase. In this paper, we report observations of another non-eruptive solar flare that exhibits an extremely large EUV late phase. A detailed data analysis shows that this large EUV late phase is mainly powered by an intense heating even earlier than the main flare heating. This paper is organized as follows. In Section 2, we describe the evolution of the flare. In Section 3, we explore the magnetic topology of the flare-hosting AR based on a non-linear force-free field (NLFFF) extrapolation, and in Section 4, we use the EBTEL model to synthesize the EUV emissions from a late-phase loop. The results are discussed in Section 5, and a brief summary is presented in Section 6.

\section{Evolution of the Flare}
\begin{figure}
\epsscale{1}
\plotone{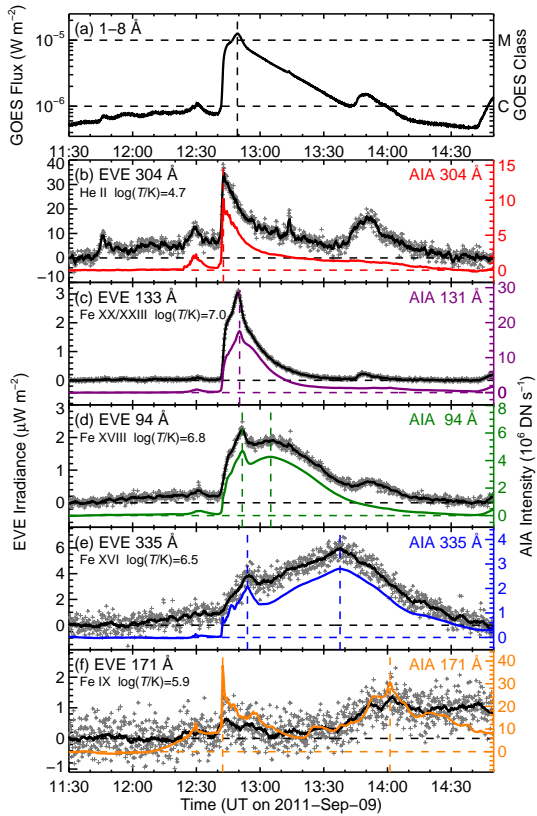}
\caption{Time profiles of the \emph{GOES} 1--8 {\AA} SXR flux (a), and the EVE full-disk irradiance (black) and AIA sub-region intensities (colored)  in several lines and passbands (b)--(f) for the 2011 September 9 M1.2 flare. The vertical dashed lines mark the peak times of the \emph{GOES} flux and AIA intensities. In deriving the EVE profiles, a line-dependent smoothing boxcar ranging from 60 to 300 s is applied to the original data points (plus signs) to enhance the ratio of signal to noise. The region over which AIA pixel count rates are summed covers a field-of-view (FOV) of 200\arcsec$\times$200$\arcsec$ enclosing the flare-hosting AR\@. Note that the EVE and AIA profiles are plotted with the corresponding background levels subtracted. For clarity, in each panel the background levels for the EVE and AIA profiles are plotted (horizontal dashed lines) with a vertical offset.}
\end{figure}

The flare-hosting AR NOAA 11283 has been extensively studied in the literature \citep[e.g.,][]{Feng13,Jiang14,LiuC14,Xu14,Romano15,Ruan15,ZhangQM15}. Its passage through the visible disk from 2011 August 31 to September 11 can be divided into two periods of roughly equal durations. During the first period, the AR was simply bipolar, indicative of a low flare level. As revealed by the \emph{GOES} flare catalog,  the AR had just harbored eight flares of classes at most C2.2 during this period. From September 4 to 5, a persistent emergence of positive magnetic flux inside the leading negative polarity of the AR was observed, which not only complicated the magnetic topology of the AR, but also enhanced its flare-productivity. As a result, a series of much more energetic flares including two X-class and five M-class ones were produced in the AR during the second period, and furthermore, some of them exhibited an evident EUV late phase. The production of EUV late phase in the X2.1 eruptive flare on  September 6 has been investigated by \citet{Dai13}.  In this work, we focus on an M1.2 non-eruptive flare on September 9 when the AR had been decaying.

\subsection{Extremely Large EUV Late Phase}
Figure~1(a) shows the \emph{GOES} 1--8~{\AA} light curve of the flare under study. Following a small bump around 12:30~UT, the SXR flux starts to rise rapidly from $\sim$12:39~UT and reaches its peak at 12:49:19~UT (outlined by the vertical dashed line). It is clearly seen that the rise phase of SXR exhibits a two-stage evolution: a very prompt rise followed by a less impulsive one. After the peak, the SXR emission also reveals two distinct decay periods:  a rapid decay, and a following much longer and less sloped one lasting until $\sim$13:40~UT when the SXR emission is elevated again. 

To further study the thermal evolution of the flare, we plot in Figures~1(b)--(f) the time profiles of the background-subtracted irradiance in several EVE spectral lines (we adopt a 5-minute average between 11:30 and 11:35~UT as the pre-flare background).  EVE measures full-disk integrated solar EUV irradiance from 1 to 1050~{\AA} with 1~{\AA} spectral resolution and 10 s time cadence. By integrating EVE spectra (EVS files) over specified spectral windows, the irradiance of some ``isolated" EUV lines (EVL files) can be derived.  Here we choose five lines from the Multiple EUV Grating Spectrographs (MEGS)-A component of EVE, which covers a wavelength range of 65--370~{\AA} with a nearly 100\% duty cycle. 

Several emission patterns of the flare can be seen from the EVE profiles. First, the cold chromospheric EVE 304~{\AA} emission (dominated by the \ion{He}{2} $\lambda$303.8~{\AA} line, $\log(T/\mathrm{K})\sim4.7$, Figure~1(b)) shows a very impulsive rise that quickly reaches its peak at 12:42:54~UT, characterizing the flare's impulsive phase. In spite of a high noise level in the original data, the smoothed profile (after applying a 300 s smoothing boxcar) of the cool coronal EVE 171~{\AA} emission (\ion{Fe}{9} $\lambda$171.1~{\AA}, $\log(T/\mathrm{K})\sim5.9$, Figure~1(f)) also reveals an obvious rise with a local maximum near the peak time in EVE 304~{\AA}. This implies that the chromosphere and transition region (TR), as responding to the initial energy release, are instantly heated to at least $\sim1$~MK \citep{Chamberlin12,Milligan12}. 

Second, the time profile of the hot coronal EVE 133~{\AA} emission (blended by \ion{Fe}{20} $\lambda$132.8~{\AA} and \ion{Fe}{23} $\lambda$132.9~{\AA}, $\log(T/\mathrm{K})\sim7$, Figure~1(c)), closely resembling the \emph{GOES} SXR light curve thanks to their similar temperature sensitivity, also exhibits a dual-rise followed by a dual-decay, and peaks nearly simultaneously with SXR at 12:49:24~UT\@. It is noted that the first-stage rise of the EVE 133~{\AA}/\emph{GOES} SXR emission is temporally correlated to the impulsive rise of the EVE 304~{\AA} emission. After the EVE 133~{\AA}  peak, the cooler coronal emissions peak sequentially at 12:51:34~UT in EVE 94~{\AA} (\ion{Fe}{18} $\lambda$93.9~{\AA}, $\log(T/\mathrm{K})\sim6.8$, Figure~1(d)) and at 12:54:44~UT in EVE 335~{\AA}  (\ion{Fe}{16} $\lambda$335.4~{\AA}, $\log(T/\mathrm{K})\sim6.5$, Figure~1(e)). The small time lags between these emission peaks indicate a fast cooling process in the corresponding flare plasma, whose inferred cooling rate ($\sim2\times10^{4}$ K s$^{-1}$) is comparable to those found in \citet{Chamberlin12}.

In addition to the first peaks, there exist second conspicuous emission peaks in both EVE 94~{\AA} and EVE 335~{\AA}, which occur at 13:05:54 and 13:37:34~UT, respectively. The time lag between the second and first peaks in EVE 335~{\AA} is 43 minutes, lying close to the lower limit of the time delay of a warm coronal late-phase peak with respect to the corresponding main flare peak (41 to 204 minutes) in the sample of \citet{Woods11}. The  emission ratio between the two peaks is $1.54\pm0.20$ (considering the fluctuations in the original data), nearly twice the average value (0.8) in \citet{Woods11}. Based on this high peak emission ratio, we adopt the same definition as that of \citet{LiuK15} and hence name this second peak an extremely large EUV late-phase peak. It is interesting that a large late phase is also evident in EVE 171~{\AA}, whose peak occurs even later (after 14:00~UT).

\subsection{Correspondence in Spatially Resolved Observations}
\begin{figure*}
\epsscale{0.9}
\plotone{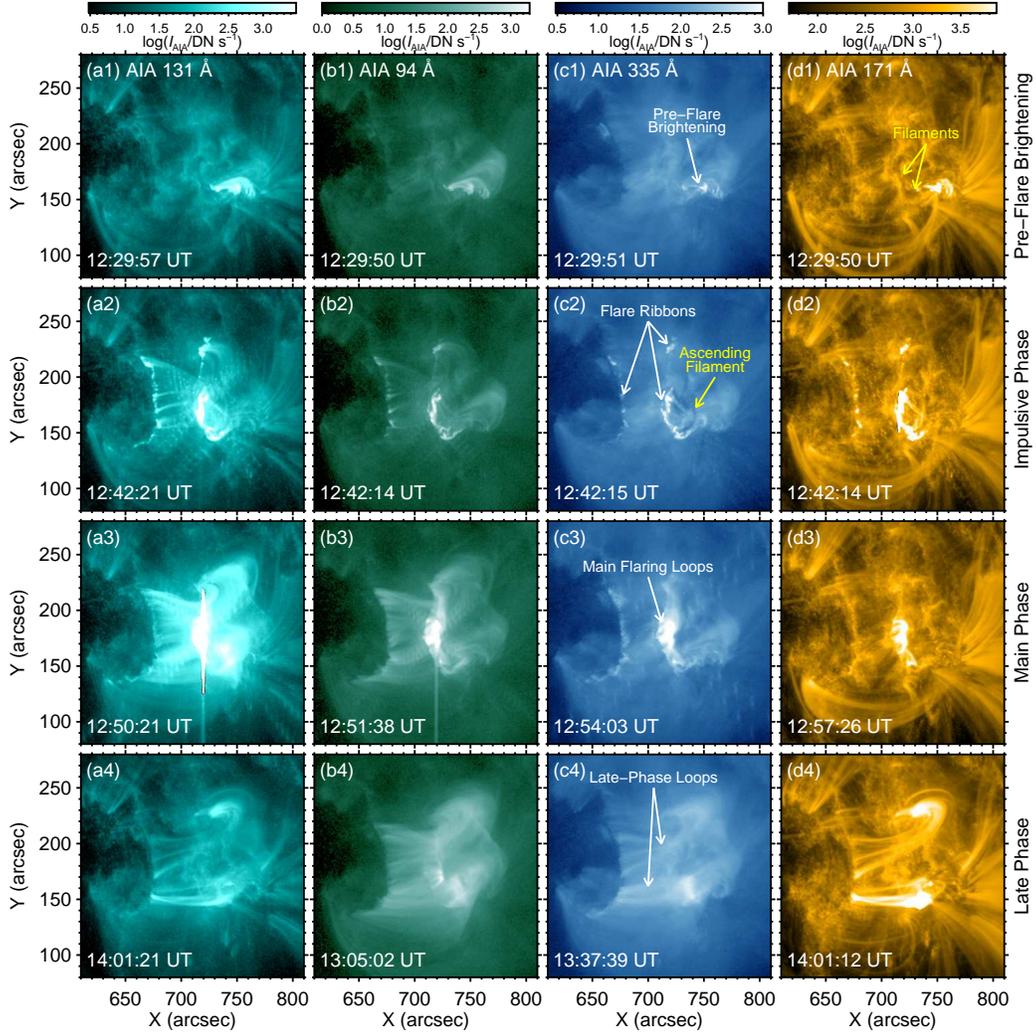}
\caption{Snapshots of the flare evolution in four AIA coronal passbands. The frames are extracted from the online animation to characterize different phases of the flare evolution, and the arrows highlight some characteristic structures, which are described in detail in the text.}
\end{figure*}

To resolve activities responsible for the flare emissions, we inspect AIA images of the flare-hosting AR\@. AIA presents multiple simultaneous full-disk images of the TR and corona in 10 passbands with a pixel size of 0$\farcs$6 and a temporal resolution of 12~s (24~s) for the EUV (UV) passbands. Here we choose a sub-region that covers a field of view (FOV) of 200\arcsec$\times$200$\arcsec$ ($x\in[610\arcsec,810\arcsec]$, $y\in[80\arcsec,280\arcsec]$) enclosing the entire AR\@. The online Animation~1 presents the flare evolution in six coronal passbands of AIA\@. Since the evolution behavior of the flare in the AIA 211 and 193~{\AA} passbands is generally similar to that in AIA 171~{\AA}, we only extract characteristic snapshots in AIA 131 (\ion{Fe}{21}, \ion{Fe}{8}), 94 (\ion{Fe}{18}, \ion{Fe}{10}), 335 (\ion{Fe}{16}), and 171~{\AA} (\ion{Fe}{9}) from the animation, and display them in Figure~2. Note that the dominant ion of each AIA passband is either the same as or of a similar formation temperature to that of the corresponding EVE line selected in Figure 1 \citep{ODwyer10}. As indicated by the arrows in Figure~2(d1), there are two low-lying filaments located in the pre-flare AR: one north-south oriented (hereafter NS), and the other east-west oriented (hereafter EW). The filaments are initially most discernible in the cool passbands such as AIA 171~{\AA} (and also AIA 131~{\AA} because of its low temperature sensitivity around 0.6 MK in addition to the high temperature response peak over 10~MK). It seems that the southern end of the NS filament is connected to the eastern end of the EW filament, apparently constituting a single L-shaped filament. Nevertheless, the following flare evolution shows that these two filaments are indeed evolving separately.

The precursor of the flare is seen as a pre-flare brightening in all AIA passbands around 12:30~UT (the first row of Figure~2), which should be responsible for the co-temporal small bumps seen in the \emph{GOES} and EVE time profiles in Figure~1. The brightening first appears in a set of sheared arcades enveloping the EW filament (highlighted by the arrow in Figure~2(c1)), and then shows a quick eastward propagation along the EW filament. When the pre-flare brightening propagates to the NS filament, this filament is destabilized and starts to ascend rapidly. As the filament rises up, its body absorbs more background coronal emission along the line-of-sight (LOS). As a result, the ascending filament becomes much more clearly seen as a curved dark feature in AIA 335~{\AA} (indicated by the yellow arrow in Figure~2(c2)). Meanwhile, multiple flare ribbons brighten up quickly in all AIA passbands, coinciding with the flare's impulsive phase (the second row of Figure~2). As pointed out by the white arrows in Figure~2(c2), the flare ribbons not only develop from two footpoints of the ascending filament, but also appear in some other remote locations to the east and north of the filament. Near the peak of the impulsive phase ($\sim$12:42~UT), several long brightening loops anchored on the eastern remote flare ribbon are already observable in the hot passbands such as AIA 131 (Figure~2(a2)) and 94~{\AA} (Figure~2(b2)), whereas they are invisible in the cooler AIA 335 (Figure~2(c2)) and 171~{\AA} (Figure~2(d2)) passbands.

The fast rising filament then slows down and finally stops at a certain altitude. Material of the filament falls down toward its two footpoints along the spine, making the filament eventually drained off (evident in Animation~1). Coronagraph images from the Large Angle and Spectrometric Coronagraph \citep[LASCO;][]{Brueckner95} on board the \emph{Solar and Heliospheric Observatory} (\emph{SOHO}) spacecraft have not recorded any coronal mass ejections (CMEs) during this period, further confirming that the flare is non-eruptive. Around the peak time of \emph{GOES} SXR emission (the third row of Figure~2), intensely brightening loops appear underneath the previously ascending filament (outlined by the arrow in Figure~2(c3)). In the framework of the CSHKP model, these loops correspond to the post-flare loops in the wake of an erupting flux rope, which exhibit the main phase of the flare. Note that in this row, each frame is taken at the time when the corresponding sub-region intensity attains a local maximum. Hence the time lags between the frames in the same row should imply a cooling process in the main flaring loops. In addition, the long brightening loops anchored on the remote flare ribbon, which can already be seen in the impulsive phase, are now fully developed as revealed in AIA 131~{\AA} (Figure~2(a3)). Compared to the main flaring loops, these longer brightening loops are less bright but cover a much larger spatial extent. Nevertheless, they are still invisible in AIA 335 (Figure~2(c3)) and 171~{\AA} (Figure~2(d3)).

The main flaring loops quickly fade out, while the long brightening loops continue to brighten up and reach an emission peak sequentially in cooler AIA passbands, as shown in the bottom row of Figure~2. Obviously, the long brightening loops in AIA 335~{\AA} (indicated by the arrows in Figure~2(c4)) are responsible for the extremely large EUV late phase revealed in EVE 335~{\AA}. Hence we are convinced to identify them as late-phase loops. Furthermore, the late-phase loops in AIA 335~{\AA} ($\sim$13:37~UT, Figure~2(c4)) are morphologically similar to those seen earlier in the hotter passbands like AIA 131 ($\sim$12:50~UT, Figure~2(a3)) and 94~{\AA} ($\sim$13:05~UT, Figure~2(b4)), and later in the cooler passband like AIA 171~{\AA} ($\sim$14:01~UT Figure~2(d4)), indicating  that they are the same structures seen in different passbands and also at different times. Here the larger time lags between the appearance times of loops in different AIA passbands reflect a much slower cooling process in the long late-phase loops than that in the main flaring loops. As mentioned above, the re-appearance of the late-phase loops in AIA 131~{\AA} around 14:01~UT (Figure~2(a4)) means that the late-phase loops have cooled down to a low temperature ($\sim$0.6~MK) to which the  AIA 131~{\AA}  passband is also sensitive.  

We further calculate the sub-region intensities by summing the count rates over all AIA pixels in the FOV of Figure~2. The intensity profiles (after subtracting a pre-flare background averaged from 11:30 to 11:35~UT, the same as that chosen in generating the EVE profiles) in five AIA passbands (here we include AIA 304~{\AA} in addition to the four passbands displayed in Figure~2) are accordingly overplotted in Figures~1(b)--(f). In each panel, the evolution of the AIA sub-region profile  is similar to that of the EVE profile, with the AIA peaks (outlined by the vertical dashed lines) occurring nearly simultaneously with the corresponding EVE peaks (typically within 1 minute). In particular for the pairs of EVE 133~{\AA}/AIA~131 {\AA}, EVE~94{\AA}/AIA 94~{\AA}, and EVE 335~{\AA}/AIA~335 {\AA}, the Pearson correlation coefficient between the two paired profiles is as high as 0.98--0.99. Moreover, the emission ratio of the late-phase peak to the main flare peak in AIA 335~{\AA} is 1.36, which is, if considering the commonly broader temperature coverage of an AIA passband compared to that of the corresponding EVE line, consistent with the ratio revealed in EVE 335~{\AA}. We have also changed the size of the sub-region used for intensity calculation. It is found that as long as the size of the sub-region is large enough to contain the whole AR, its variation only marginally affects the resulting AIA profiles. All these factors suggest that during this period, the activities taking place inside the AR contribute predominantly to the variabilities in full-disk integrated solar emissions in EUV wavelengths, especially for those formed in hot and warm coronal plasmas. Since the \ion{Fe}{9} emission mainly comes from the cool bulk corona, emission fluctuations outside the AR may significantly influence the overall irradiance in EVE 171~{\AA}, hence making the original data for this line rather noisy, as opposed to those in other EVE lines. Finally, it is worth noting that there are a small spike around 13:15~UT and a moderate bump after 13:40~UT in the \emph{GOES} and EVE profiles, which nevertheless have no counterparts in the AIA sub-region profiles. By carefully checking the AIA full-disk images, it is found that they are caused by two small brightening events occurring in another AR NOAA 11289. 

\subsection{Two-Stage Energy Release}
\begin{figure*}
\epsscale{1}
\plotone{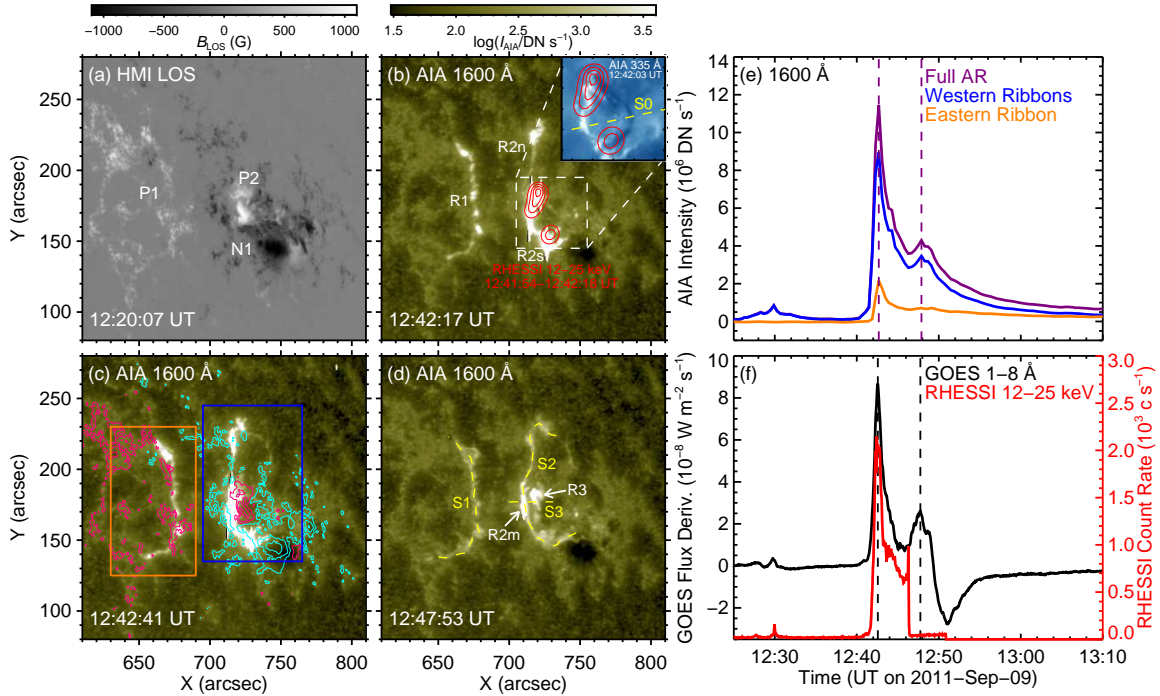}
\caption{Left and middle: HMI LOS magnetogram of AR 11283 taken at 12:20~UT (a), and snapshots showing the flare evolution in AIA 1600~{\AA} (b)--(d), which are extracted from the online animation. The LOS magnetogram reveals three main magnetic polarities, whose contours are overplotted in panel (c), with levels of $\pm200$, $\pm600$, and $\pm1000$~G (pink for positive, and cyan for negative). Multiple flare ribbons (prefixed with the letter ``R") are identified in the AIA 1600~{\AA} images. The inset in panel (b) is a simultaneous AIA 335~{\AA} image, whose FOV is outlined by the dashed box. Also overplotted in this panel is a simultaneous \emph{RHESSI} image at energies of 12--25~keV reconstructed with the Pixon algorithm, with contour levels corresponding to 30\%, 50\%, 70\%, and 90\% of the maximum intensity, respectively. Four slices (S0--S3) are drawn as the yellow dashed lines or curves in panels (b) and (d), for which the time-distance plots are presented in Figure 4. Right: time profile of the AIA 1600~{\AA} intensity for different regions (e) and time profiles of soft (derivative) and HXRs (f). The vertical dashed lines indicate the peak times for the corresponding profiles. The AIA intensity is integrated over sub-regions enclosed by the colored boxes in panel (c).}
\end{figure*}

The evolution of the flare is also traced in the chromospheric AIA 1600~{\AA} passband, as demonstrated in the online Animation~2. Emission in this passband includes the continuum formed around the temperature minimum region and the \ion{C}{4} $\lambda$1548.2/1550.8~{\AA} doublet lines formed in the upper chromosphere and TR\@. During a solar flare, the optically thin \ion{C}{4} lines are significantly enhanced, showing a prompt and sensitive response to the flare energy release. Therefore, the evolution in AIA 1600~{\AA} can be regarded as a proxy for flare heating \citep{LiY12,Qiu12,Zhu18}. Some snapshots of the animation, as well as an LOS magnetogram of the AR taken near the time of the flare by the Helioseismic and Magnetic Imager \citep[HMI;][]{Scherrer12} on board \emph{SDO}, are displayed in Figures~3(a)--(d). Three main magnetic polarities are identified  and labeled as P1, N1, and P2, of which P2 is a parasitic positive polarity embedded in the host negative polarity N1 (Figure~3(a)). During the course of the flare, flare ribbons are observed in all of these polarities (Figures~3(b)--(d)), like those seen in Figure~2 but free of any possible contaminations from the loop structures. The appearance of multiple flare ribbons is believed to be a natural consequence of the multipolar magnetic field. When overplotting the contours of the LOS magnetic field onto the AIA 1600~{\AA} image (Figure~3(c)), it is seen that the flare ribbons generally occur in locations of magnetic concentration.

The AIA 1600~{\AA} intensity profile of the AR (purple) shown in Figure~3(e) reveals two peaks at 12:42:41 and 12:47:53~UT (highlighted by the vertical dashed lines), respectively, indicative of two main episodes of energy release taking place in quick succession. With the aid of the profile, we select the AIA 1600~{\AA} snapshots close to or at the times of the intensity peaks, as displayed in Figures~3(b)--(d). The first-stage energy release is very energetic; flare ribbons brighten up promptly, coinciding with the flare's impulsive phase. Around the peak of the ribbon brightening (Figures~3(b) and (c)), the flare ribbons are manifested as an elongated one (R1) located in the eastern positive polarity P1 and a semi-circular one (R2) located in the western negative polarity N1. R1 corresponds to the remote flare ribbon seen in Figure~2, whereas R2 is not continuous but broken at its middle into two segments: R2n in the north and R2s in the south. A third flare ribbon is believed to exist in the parasitic positive polarity P2. However, saturation of the CCD pixels in this region largely masks the rather compact ribbon, making it indistinguishable from the nearby ribbon segment R2s. To validate our speculation, we reconstruct the \emph{Reuven Ramaty High Energy Solar Spectroscopic Imager} \citep[\emph{RHESSI};][]{LinRP02} image at the energy band of 12--25~keV in the interval of 12:41:54--12:42:18~UT, which reveals two HXR sources (outlined by the red contours in Figure~3(b)) located in the ``masked" compact ribbon and the southern ribbon segment R2s, respectively. When overplotting the HXR sources onto the simultaneous AIA 335~{\AA} image inserted in Figure~3(b), it is further found that they are excellently co-spatial with the two footpoints of the ascending NS filament.

Compared to the first-stage energy release, the second-stage energy release is more gentle; the flare ribbons re-brighten up moderately after the first peak. For the eastern remote ribbon R1, the brightening is observed to propagate from north to south along the ribbon. For the western semi-circular ribbon R2, a newly brightening ribbon segment R2m appears and becomes the most prominent section along this ribbon, filling the gap between the previously disconnected ribbon segments R2n and R2s. Meanwhile, the previously masked compact ribbon in the parasitic polarity P2 now becomes clearly discernible, which is identified as R3 (Figure~3(d)). As the energy release goes on, both R2m and R3 experience a separation motion away from the polarity inversion line (PIL) between them, which is usually observed in a two-ribbon solar flare. All these evolution patterns are clearly revealed in Animation 2. 

In addition to the time profile of the full AR, in Figure~3(e) we also plot the time profiles of the AIA 1600~{\AA} intensity for regions covering the eastern and western ribbons, which are outlined in Figure~3(c) by the orange and blue boxes, respectively. During the first-stage energy release, the intensities in both regions show a synchronous enhancement, with the first peaks in all time profiles occurring at exactly the same time. Nevertheless, the second intensity  peak for the region of eastern ribbon is much less discernible than that for the region of  western ribbons. It implies that during the second-stage energy release, the majority of the released energy should be deposited into the loops anchored on the western ribbons. 

\begin{figure}
\epsscale{1}
\plotone{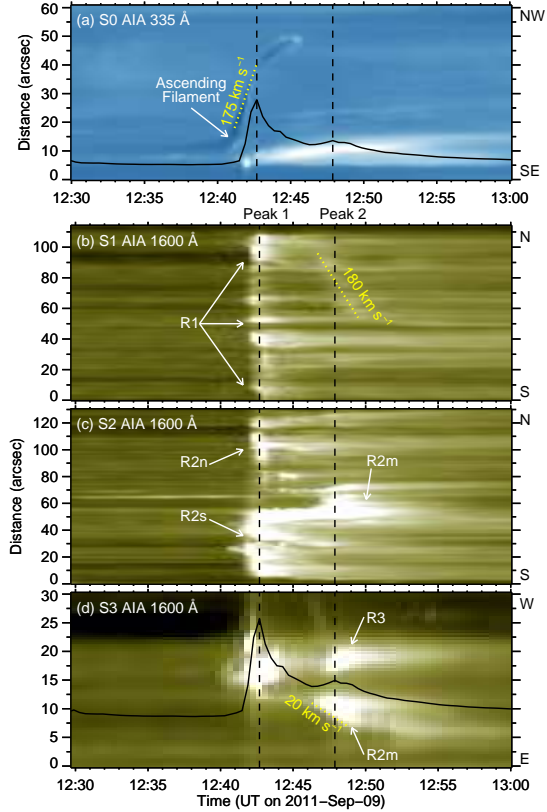}
\caption{Time-distance plots along the selected slices in AIA 335 {\AA} (a) and 1600 {\AA} (b)--(d), within which some characteristic structures are identified and labeled. Linear fittings are applied to some moving structures traced by the yellow dotted lines, with the fitted speeds labeled along these lines. Overplotted in panels (a) and (d) is the AIA 1600 {\AA} intensity profile of the AR (the same as that plotted in Figure~3(e)) revealing two peaks  by the vertical dashed lines in all panels.}
\end{figure}

We use time--distance plots to further study the energy release process. Four slices are selected and drawn in Figures~3(b) and (d), among which S0 extends across the apex of the ascending filament while S1--S3 lie either parallelly or perpendicularly to the flare ribbons. The temporal evolution along  these slices in  AIA 335~{\AA} (S0) and 1600~{\AA} (S1--S3) is shown in Figure~4. A close correlation between the rise of the filament and the brightening of the flare ribbons is found. A linear fit to the trajectory of the ascending filament apex reveals a rise speed of 175~km~s$^{-1}$ projected onto the plane of the sky (Figure~4(a)). Coinciding with the fast rise of the filament, the flare ribbons (R1, R2n, R2s, and possibly masked R3) brighten up simultaneously and quickly reach the first peaks exactly at the time when the fast ascending filament turns to decelerate (Figures~4(b)--(d)). As its ascending speed gradually approaches zero, the filament  disappears abruptly in the S0 plot, which is due to the falling of material from the filament apex. 

Shortly after the ``disappearance" of the filament (within 1 minute), the second-stage energy release comes into play. Brightening occurs successively from north to south along the northern section of R1, revealing an apparent motion speed of 180~km~s$^{-1}$ (Figure~4(b)). Elongation motions of flare ribbon brightening along the PIL have been both theoretically modeled and observationally seen in solar flares \citep[e.g.,][]{Priest17,Qiu17}. In this case, the observed slipping of the ribbon brightening is more likely  linked to 3D slipping/slip-running reconnection \citep{Aulanier06,Aulanier07} taking place in a quasi-separatrix layer \citep[QSL;][]{Demoulin96}. In addition to the apparent elongation motion of R1, both the newly brightening ribbon segment R2m (Figure~4(c)) and the previously masked ribbon R3 show a separation motion from the PIL (Figure~4(d)). Compared to R3, the motion of R2m is more reliable to trace. A linear fit to the outer boundary of R2m reveals a separation speed of 20~km~s$^{-1}$. To compensate for the projection shortening  at the flare position of $\sim$N13$\degr$W52$\degr$, we multiply the fitted speed by a factor of $\sim$1.6, and obtain a corrected separation speed of $\sim$30~km~s$^{-1}$, which is consistent with the results previously found in two-ribbon flares \citep[e.g.,][]{Asai04,Miklenic07}.

The two episodes of energy release are also evident in X-rays. Figure~3(f) shows the derivative of the \emph{GOES} 1--8~{\AA} flux (black), which is generally taken as a proxy for the HXR emission owing to the ``Neupert effect" \citep{Neupert68}. Two peaks are readily resolved at 12:42:33 and 12:47:45~UT (indicated by the vertical dashed lines), respectively, in perfect coincidence with the corresponding AIA 1600~{\AA} peaks (within 10~s). The overplotted \emph{RHESSI} 12--25~keV light curve (red) reveals only one peak at 12:42:26~UT, which coincides with the first peak in the \emph{GOES} derivative curve. The reason for the absence of a second peak in the \emph{RHESSI} HXR light curve is that the satellite has orbited into the Earth shade at this time.  

\section{Magnetic Topology of the AR}
\begin{figure*}
\epsscale{0.9}
\plotone{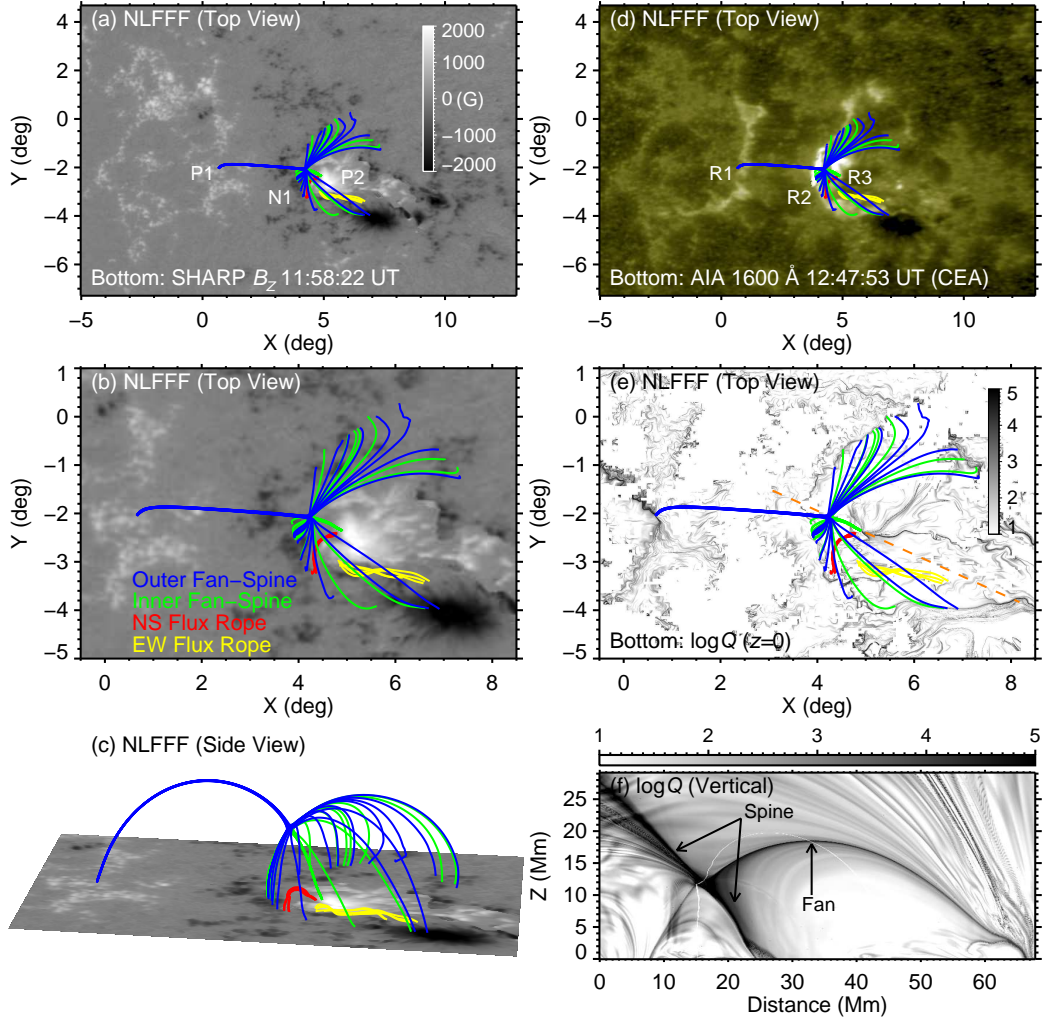}
\caption{NLFFF extrapolated magnetic field lines showing fan-spine and flux rope structures, which are overlaid on the boundary SHARP $B_z$ magnetogram (a)--(c), an AIA 1600~{\AA} CEA map (d), and a filtered map of the squashing factor $Q$ on the bottom boundary (e), respectively. Also shown is a vertical $Q$-map revealing a fan-spine QSL structure (f). The field lines are color-coded according to the legends in panel (b), and the $Q$-map is shown for the vertical cut along the orange dashed line plotted in panel (e). In the left column, panels (b) and (c) show zoomed-in top and side views of the field lines, respectively. Note that the coordinate origin in the top-view panels corresponds to the projection center of the original SHARP CEA map, which has a Carrington coordinate of (220.28$\degr$, 17.39$\degr$).}
\end{figure*}

Magnetic topology is crucial for understanding the magnetic reconnection process in a solar flare and the consequent morphological evolution of the flare loops and ribbons. In a typically low $\beta$ (the ratio of gas pressure to magnetic pressure) corona, we explore the magnetic topology of the flare-hosting AR with the aid of an NLFFF extrapolation.

We adopt the optimization method proposed by \citet{Wheatland00} and implemented by \citet{Wiegelmann04} to perform the NLFFF extrapolation. To  prepare the bottom boundary, we use an HMI vector magnetogram of the AR taken close to the time of flare occurrence ($\sim$11:58~UT), which is selected  from the Space-weather HMI Active Region Patches (SHARPs). The pipeline to generate SHARP data products includes the removal of the 180$\degr$ ambiguity in transverse field \citep{Metcalf06,Leka09}, and the remapping of the magnetic field vector \citep{Gary90} in a cylindrical equal-area \citep[CEA;][]{Calabretta02} projection (for more details, see \citeauthor{Bobra14} \citeyear{Bobra14} and \citeauthor{Hoeksema14} \citeyear{Hoeksema14}). The grid size used for the SHARP sampling is 0.03$\degr$ in heliocentric angle, corresponding to a pixel size of $\sim$0.36~Mm. To save the computation time while not significantly losing the essential features in the magnetic field modeling, we extract a sub-region that is large enough to cover all main polarities in the original SHARP magnetogram, and degrade the sampling resolution by a factor of 2. Further preprocessing is applied to the area of interest to minimize the net magnetic force and torque on the boundary \citep{Wiegelmann06}. The extrapolation is then conducted in a box of $300\times200\times200$ grid points uniformly spaced in the computation domain. 

The magnetic topology of the AR based on our NLFFF extrapolation is presented in Figure~5. The most prominent features are a 3D magnetic null point located at a height of 11.8~Mm, and inner (green) and outer (blue) fan-spine magnetic field lines passing around it \citep{Lau90}. With the aid of the normal component ($B_z$) of the SHARP vector magnetogram (Figures~5(a)--(c)), it is seen that the footpoints of the inner and outer spine field lines reside in the parasitic and eastern positive polarities P2 and P1, respectively, while the fan field lines form a dome-like structure, whose base intersects with the solar surface in the negative polarity N1. Embedded in the fan dome, two magnetic flux ropes (red for the NS-oriented one and yellow for the EW-oriented one) are found to exist. Through co-alignment with the AIA coronal images (not shown here), it is found that the flux ropes are approximately co-spatial with the two low-lying filaments identified before the flare.

In Figure~5(d) we overlay the above magnetic field lines on an AIA 1600~{\AA} image taken at the second peak in this passband (12:47:53~UT) when all the flare ribbons are clearly seen. For a better comparison between them, we remap the AIA 1600~{\AA} image into the same CEA projection as that of the SHARP map. Note that such remapping will introduce artificial distortions to above-the-surface structures, especially for those located far away from the disk center, just like the late-phase loops observed in this flare. For this reason, we do not attempt to compare the extrapolated magnetic field with the flare loops seen in other AIA coronal passbands. As revealed in Figure~5(d), there is a good match between the footpoints of the overlaid magnetic field lines and the chromospheric flare ribbons. The footpoints of the inner and outer spines are located exactly in the compact ribbon R3 and the eastern remote ribbon R1, respectively. The degree of match found here seems better than those in previously reported cases where the spatial discrepancy between the outer spine footpoint and the remote flare ribbon is typically over 10~Mm \citep[e.g.,][]{Vemareddy14,YangK15,Masson17}. Except for the northernmost section, the circular ribbon R2 also matches the base of the fan dome very well. In addition, the two footpoints of the NS flux rope are anchored at ribbons R2 and R3, respectively. 

The null point and the fan-spine structure indicate the presence of magnetic separatrix  and QSLs, whose distribution can be quantified by the so-called squashing factor $Q$ \citep{Titov02}. The $Q$ factor quantitatively characterizes the connectivity gradients between one magnetic field line and its neighboring field lines. Here we use the method proposed by \citet{Pariat12} to compute the distribution of $Q$ value in the 3D domain. In Figure~5(e) we plot the distribution of $\log Q$ on the bottom boundary ($z=0$). For the sake of clarity, we apply a filter to the $Q$-map so that only the distribution of $\log Q$ in regions of $|B_z|>$100~G is displayed. When overlaying the modeled field lines on the $Q$-map, it is clearly seen that the fan-spine field lines intersect with the solar surface at patches of high $\log Q$ values. This is consistent with the observational fact found in many previous studies \citep[e.g.,][]{ChenPF12,Masson17} that flare ribbons typically stop at the intersection of QSLs with the solar surface. To depict the structure of the QSLs in the corona, we draw a slice across the fan-dome base and the footpoint of the inner spine on the bottom boundary (the orange dashed line in Figure~5(e)), and plot the 2D $Q$-map in the vertical plane extending above the slice, which is shown in Figure~5(f). As expected, it clearly shows the fan-spine magnetic topology of this event, i.e., a QSL surrounding the fan-spine separatrix \citep{Masson09}.

There are still some discrepancies in morphology between the extrapolated magnetic field and the flare ribbons. First, the northernmost section of the circular ribbon R2 extends well beyond the fan-dome base, showing a spatial offset up to 15~Mm (Figure~5(d)). Second, in spite of the excellent match between the outer spine footpoint and the eastern remote ribbon R1, field lines that originate from other parts of this elongated ribbon, pass through the fan-spine QSL, and end up at the circular ribbon R2 cannot be reasonably figured out, although their existence can be confirmed by the appearance of the  widespread late-phase loops. For this reason, we do not display these field lines in Figure~5. As mentioned above, at the time of the flare, the AR is located far away from the disk center. The SHARP remapping procedure could hence amplify the errors of magnetic field measured in the local plane, which are then transferred to all the field components at the bottom boundary used for extrapolation. This may significantly compromise the fidelity of magnetic field extrapolation from regions of relatively weak field strengths.

\section{Modeling of the Late-Phase Loops}
\subsection{Stereoscopic Measurement of the Loop Lengths}
\begin{figure*}
\epsscale{0.8}
\plotone{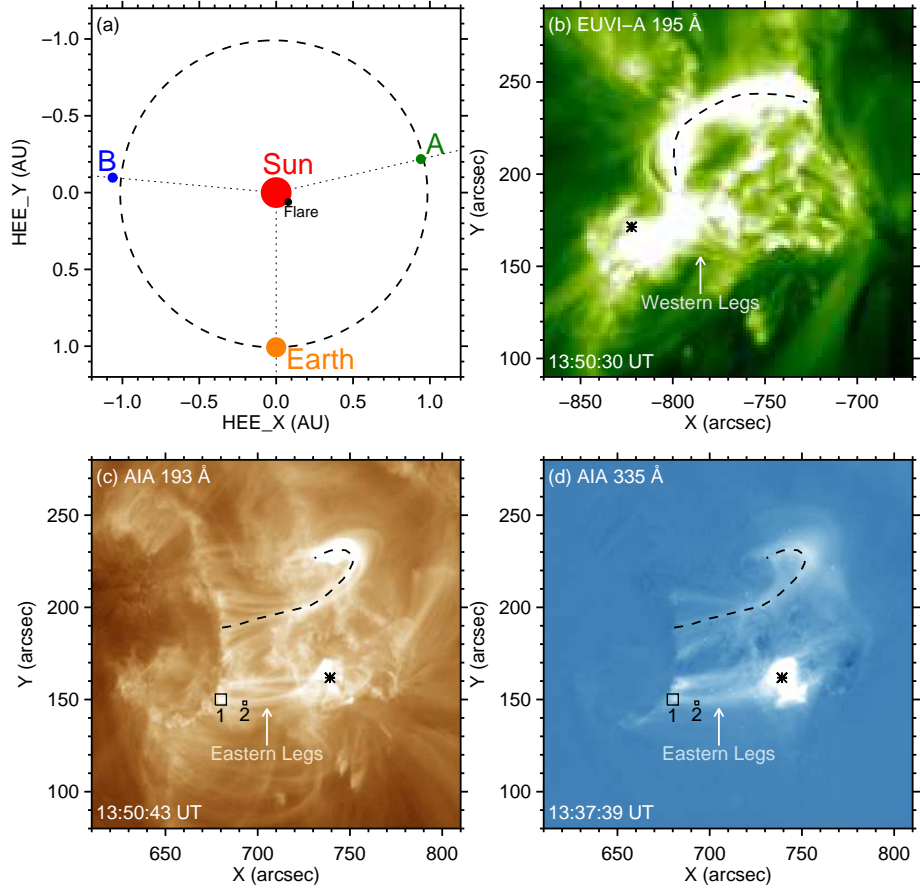}
\caption{Deployment of the \emph{STEREO} twin spacecraft with respect to the Earth at the time of the flare (a), and images of the late-phase loops viewed from two different perspectives (b)--(d). In panel (a), the black dot  indicates the longitudinal position of the flare site. In panels (b)--(d), the dashed lines and asterisks outline the projection of the characteristic late-phase loops onto the two different perspectives, respectively, and the arrows point to the legs  of the southern late-phase loops. In the bottom row images, two boxes are drawn at the leg and footpoint of a southern late-phase loop, respectively, over which we derive the AIA light curves used for modeling of the late-phase emissions.}
\end{figure*}

By combing the full-disk integrated EVE and spatially resolved AIA observations, a clear correspondence between the late-phase emissions and the long late-phase loops in this flare has been validated. The length of a coronal loop is, both theoretically and observationally, an important parameter for the loop evolution. In previous case studies, estimation of the lengths of late-phase loops was mainly based on single-perspective observations and therefore subject to large uncertainties. Through imaging observations from multiple perspectives, we could in principle track the realistic coronal loops in 3D, and hence more accurately determine the loop lengths. 

Figure~6(a) depicts the deployment of the \emph{Solar Terrestrial Relations Observatory} \citep[\emph{STEREO};][]{Kaiser08} twin spacecraft  with respect to the Earth at the time of the flare. The longitudinal position of the flare site (indicated by the black dot) and its favorable viewing angles to \emph{STEREO-A} and near-Earth \emph{SDO} enable us to stereoscopically measure the 3D coordinates of flare loops captured by both satellites. In Figures~6(b) and (c) we display paired images of the late-phase loops taken simultaneously at $\sim$13:50~UT by the EUV Imager \citep[EUVI;][]{Wuelser04} on board \emph{STEREO-A} (hereafter EUVI-A) in 195~{\AA} and by AIA in 193~{\AA}, respectively, both of which have a similar temperature response peaking around 1.3~MK\@. As shown in the figures, the late-phase loops appear mainly in two clusters. In the north, the late-phase loops are readily resolved in both passbands. Using the Solar Software \citep[SSW;][]{Freeland98} routine \texttt{scc\_measure.pro}, we trace a full late-phase loop, whose 3D coordinates are projected onto the two perspectives (dashed lines), respectively. By integrating the distance along the loop, we derive a length of 81~Mm for this late-phase loop. In the south, the triangulation measurement of the late-phase loops is nevertheless significantly affected by an LOS effect.  When viewed from the two different perspectives, either the eastern (from EUVI-A) or the western (from AIA) legs of the late-phase loops coincidentally overlap along the corresponding LOS\@. As a result, only the legs on the opposite side are distinguishable from the corresponding perspective, as pointed out by the arrows in the figures. Moreover, the intensity of the loop-top regions in both passbands is predominately high, since the intensity observed there is actually an integration along the LOS, which is also contributed by emissions from the loop legs. Restricted by the LOS effect, here we just stereoscopically estimate the apex location of the late-phase loops (indicated by the asterisks), whose height is $\sim$35~Mm above the solar surface. Assuming a semi-circular loop geometry, we obtain a loop length of $\sim$110~Mm. Note that due to the uncertainty in determining the loop apex, this value might be an underestimation of the lengths for the southern late-phase loops.

The projections of the axis of the  late-phase loop in the north and apex of the late-phase loop in the south are further overplotted in an AIA 335~{\AA} difference image (Figure~6(d)) taken 13 minutes earlier (at the late-phase peak in AIA 335~{\AA}). We find an excellent match between the loop geometries and emission features in the two AIA passbands, further confirming the slow cooling process in the long late-phase loops.

For comparison, we also quantify the lengths of the main flaring loops, which are simply inferred from the separation distance between the flare ribbons R2m and R3 on which the two conjugate loop footpoints are anchored. The estimated lengths of the main flaring loops are 10--25~Mm. Obviously, the two sets of flare loops are distinct in length, with a length ratio between the late-phase loops and the main flaring loops exceeding 3.

\subsection{EBTEL Modeling of the Late-Phase Emissions}
The electromagnetic emissions from a flaring loop reflect the underlying evolution of temperature and density, and thereby the energy release history in the loop. Using the EBTEL model, we model the EUV emissions of a late-phase loop in this flare. By comparing the synthetic EUV light curves with the observed light curves in several AIA passbands, we quantify  parameters of the heating on the late-phase loop as well as the loop geometry.

\subsubsection{Model Setup}
The EBTEL model is a zero-dimensional (0D) hydrodynamic model that describes the evolution of the average temperature, pressure, and density along a coronal loop (or more strictly speaking, a coronal strand). The basic idea behind EBTEL is that the imbalance between the heat flux conducting down into the TR and the radiative loss rate there leads to an enthalpy flux at the coronal base: an excess heat flux drives an evaporative upflow, whereas a deficient heat flux is compensated for by a condensation downflow. This upward/downward enthalpy flow controls the hydrodynamic evolution of the coronal loop.

Another assumption of EBTEL is that any flows along the loop are subsonic, which should hold for most of the time of the loop evolution. This means that the kinetic energy is negligible compared to the internal energy. Hence, the equation of momentum can be dropped from the governing equations and the kinetic terms can be dropped from the equation of energy conservation. Following this way, in EBTEL we solve the remaining loop-averaged (along the coronal section of the loop) hydrodynamic equations, including the equations of continuity and energy conservation, and equation of state, which are given by
\begin{equation}
\frac{dn}{dt}=-\frac{c_2}{5c_3kTL}(F_0+c_1\mathcal{R}_c),
\end{equation}
\begin{equation}
\frac{dp}{dt}=\frac{2}{3}\left[H-(1+c_1)\frac{\mathcal{R}_c}{L}\right],
\end{equation}
and
\begin{equation}
\frac{1}{T}\frac{dT}{dt}=\frac{1}{p}\frac{dp}{dt}-\frac{1}{n}\frac{dn}{dt},
\end{equation}
where $n$, $p$, and $T$ are the average density, pressure, and temperature of the coronal loop, respectively, $L$ is the loop half-length (EBTEL assumes a symmetric semi-circular loop geometry and hence treats only half of the loop), $k$ is the Boltzmann constant, $c_2$ ($c_3$) is the ratio of the average coronal (coronal base) temperature to the apex temperature, $F_0=-(2/7)\kappa_0(T/c_2)^{7/2}/L$ (where $\kappa_0=8.12\times10^{-7}$ in cgs units is the classical Spitzer thermal conduction coefficient) is the heat flux at the coronal base, $\mathcal{R}_c={n}^2\Lambda(T)L$ (where $\Lambda(T)$ is the optically thin radiative loss function) approximates the radiative loss rate from the corona, $c_1=\mathcal{R}_{tr}/\mathcal{R}_c$ is the ratio of radiative loss rate of the TR to that of the corona, and $H$ denotes the average volumetric heating rate (note that here we use $H$ instead of $Q$, the commonly adopted notation for heating rate in the literature, to avoid confusion with the squashing factor introduced in Section 3).

Of the three dimensionless parameters $c_1$, $c_2$, and $c_3$ in EBTEL, $c_1$ is the most important. The main difference between the different versions of EBTEL lies in the treatment of $c_1$. Assuming a single power-law radiative loss function, \citet{Martens10} analytically studied the energy equilibrium for a static coronal loop, whose solutions reveal a value of $c_1$ of around 2, and values of $c_2$ and $c_3$ close to 0.9 and 0.6, respectively. In the original EBTEL model \citep{Klimchuk08}, to achieve an overall consistency with the 1D simulation results, $c_1$ was kept as a constant of 4 throughout the loop evolution. Nevertheless, when a loop evolves dynamically,  the equilibrium assumption is broken and $c_1$ also evolves. Obviously, the choice of a fixed $c_1$ value is not physically reasonable. In the later EBTEL versions, more physics was included to appropriately determine the values of $c_1$.  One piece of physics previously ignored is gravitational stratification, whose main effect is to depress the coronal radiation. Thus, higher values of $c_1$ can be expected, especially for long loops with large ratios of the length to the gravitational scale height \citep{Cargill12}. Another piece is the deviation of the dynamically cooling loop from equilibrium states. In the early conductive cooling phase, the density increase in response to the strong heat flux is not so fast that the loop is under-dense with respect to a static loop at the same temperature, leading to higher values of  $c_1$ \citep{Barnes16}. During the radiative cooling phase, the loop is instead over-dense, which in turn reduces the $c_1$ values \citep{Cargill12}. In our modeling, we hold $c_2$ and $c_3$  at typical values of 0.9 and 0.6, respectively, as specified in all EBTEL versions, whereas we use the latest EBTEL version \citep{Barnes16} to consistently calculate the $c_1$ values. 

\subsubsection{Comparison with Observations}
\begin{figure}
\epsscale{1}
\plotone{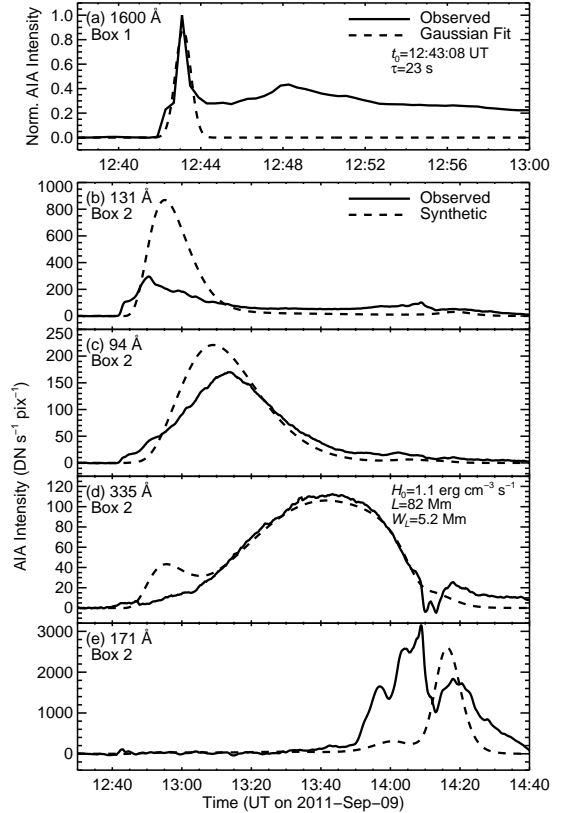}
\caption{Top: Normalized time profile  observed at the footpoint of the selected late-phase loop (Box~1 in Figure~6) in the AIA 1600~{\AA} passband (solid curve), with the Gaussian fit  overplotted (dashed curve). Bottom: Light curves observed at the leg of the same loop (Box~2 in Figure~6) in four AIA coronal passbands (solid curves), as well as the synthetic light curves (dashed curves) computed with the best-fit parameters in the EBTEL modeling (specified in the legends in panel (d)).}
\end{figure}

Based on the output temperature and density from the EBTEL model, we can synthesize the emissions from the loop. The synthetic intensity of the loop in an AIA passband can be computed as
\begin{equation}
I_{\mathrm{AIA}}=R(T)n^2W_L\ (\mathrm{DN}\ \mathrm{s}^{-1}\ \mathrm{pix}^{-1}),
\end{equation}
where $R(T)$ is the temperature response function of the AIA passband, which can be obtained by the SSW routine \texttt{aia\_get\_response.pro}, and $W_L$ is the nominal width (depth) of the loop along the LOS\@. In generating the AIA response functions, we set the \texttt{evenorm} keyword, which will give better agreement with full-disk EVE observations when cross-checking data from both AIA and EVE\@. Due to the 0D nature of the EBTEL model, it is physically not applicable to give a spatial distribution of the synthetic intensities along the loop. To compare the synthetic intensities with observations, we select a late-phase loop in the south whose observed intensities are roughly uniformly distributed along its eastern leg, as shown in the bottom row AIA images of Figure~6. We draw a $2\arcsec\times2\arcsec$ box over the loop leg (Box~2 in Figure~6). The intensities averaged over this box in several AIA coronal passbands are taken as the observed loop intensities, whose time profiles are shown as the solid curves in Figures~7(b)--(e). 

To infer the heating function for the loop, we calculate the chromospheric light curve at the loop footpoint, as proposed by \citet{LiY12}, \citet{Qiu12}, and \citet{Zhu18}. To this end, we draw a $6\arcsec\times6\arcsec$ box (Box~1 in Figure~6) at the eastern footpoint of the selected late-phase loop, which is located on the remote flare ribbon R1. In Figure~7(a) we plot the normalized AIA 1600~{\AA} intensity profile averaged over this box (the solid curve), which closely resembles the intensity profile for the whole eastern ribbon shown in Figure~3(e). By applying a Gaussian fit to the rise and fast decay phase of the AIA 1600~{\AA} light curve, we derive a peak time at 12:43:08~UT and a width of 23~s for the fitted Gaussian function (the dashed curve in Figure~7(a)). Mimicking the shape of the fitted Gaussian function, the imposed impulsive heating on the late-phase loop is then given by
\begin{equation}
H(t)=H_0\exp\left[-\frac{(t-t_0)^2}{2\tau^2}\right]\ (\mathrm{erg}\ \mathrm{cm}^{-3}\ \mathrm{s}^{-1}),
\end{equation}
where $H_0$ is the amplitude of the heating, $t_0$=12:43:08~UT the time of the peak heating rate, and $\tau$=23~s the Gaussian duration of the heating. Note that there is a second small peak $\sim$5 minutes after the first peak in the AIA 1600~{\AA} intensity profile, which possibly reflects a second heating process. Nevertheless, this second heating should be much weaker than the first heating, and hence can be safely ignored in our modeling. In addition to the impulsive heating, we also include a steady background heating whose rate is fixed to be 10$^{-5}$~erg~cm$^{-3}$~s$^{-1}$. While maintaining the loop temperature and density before and long after the impulsive heating, this background heating does not obviously affect the modeled EUV emissions from the loop.

There are three free parameters in our modeling, i.e., $H_0$, $L$, and $W_L$. Although the characteristic loop length for the southern late-phase loops has been estimated using the triangulation method, the LOS effect may introduce some uncertainties. Therefore, here we still regard $L$ as a free parameter. We use the MPFIT \citep{Markwardt09} package distributed in SSW  to perform the fitting. In our modeling, we only fit the observed light curve of the late-phase loop in AIA 335~{\AA}, since we mainly focus on the EUV late phase in the warm coronal emissions. For other passbands, we manually inspect the degree of match between the synthetic light curves computed with the best-fit parameters and the observed light curves. Some factors that could affect the loop emissions in these passbands  are addressed later in this section.

The fitting gives best-fit values for the three free parameters: $H_0$=1.1~erg~cm$^{-3}$~s$^{-1}$, $L$=82~Mm, and $W_L$=5.2~Mm. To test the robustness of the fitting, we vary the starting guesses for the parameters within a reasonable range. In most cases, the fitting procedure converges to the same results as those given above. The synthetic light curves computed with the best-fit parameters are overplotted as the dashed curves in Figures~7(b)--(e).  As seen in Figure~7(d), the peak in both the synthetic and the observed light curves in AIA 335~{\AA} occurs almost one hour after the impulsive heating. In the other passbands, the peak times in the synthetic light curves also reasonably agree with those in the observed light curves. This indicates that even without an additional delayed heating, the long cooling process of the late-phase loop can sufficiently explain the presence of the late-phase emissions at different temperatures. Note that there are three consecutive peaks in the observed AIA 171~{\AA} light curve. It means that the LOS should intersect at least three late-phase loops of similar half-lengths, and the parameter $W_L$ in the modeling would reflect a combination of the widths of all these loops seen along the LOS\@. In this sense, the effective width of an individual late-phase loop is of the order of 1.7~Mm, which is consistent with those found in previous observations \citep[e.g.,][]{Aschwanden08}.

In spite of a good match between the modeled results and observations seen in AIA 335~{\AA}, there is a  large discrepancy in amplitude between the synthetic light curves and the observed ones in the other passbands. In the hotter AIA 131 and 94~{\AA} passbands, the synthetic peak intensities are significantly higher than those observed. The hotter the passband, the more prominent such a discrepancy (Figures~7(b) and (c)). We attribute this discrepancy to a possible deviation from an ionization equilibrium in the late-phase loops, a factor not considered in our EBTEL modeling. During the impulsive heating and conductive cooling stage, the loop temperature evolves so fast that the ionization process cannot catch up with the temperature change, resulting in a non-equilibrium ionization \citep{Reale08,Bradshaw11}. Obviously, this effect is the most prominent for hot coronal emissions from long tenuous loops, e.g., the late-phase loops. Using the full 1D HYDRAD code \citep{Bradshaw03}, \citet{Bradshaw11} numerically studied the effect of non-equilibrium ionization on loop emissions, and found that the synthetic loop intensities in hot passbands such as AIA 131~{\AA} can be depressed by orders of magnitude compared to those under the ionization equilibrium assumption. They also pointed out that as the cooling rate slows down and the loop density increases, the ions can have enough time to reach an ionization equilibrium at a medium temperature like 6~MK. Therefore, emissions formed below this temperature (like the EUV late-phase emission in AIA 335~{\AA}) are unlikely to be notably affected by this effect. Another piece of evidence for the existence of non-equilibrium ionization in the modeled late-phase loop is the small bump seen in the synthetic AIA 335~{\AA} light curve shortly after the impulsive heating, which is also synchronous with the peak of the synthetic light curve in AIA 131~{\AA}\@. This bump reflects a high temperature response over 10~MK in AIA 335~{\AA}, which is introduced by crosstalk from the AIA 131~{\AA} passband \citep{Boerner12}. Nevertheless, due to the possible non-equilibrium ionization at flare temperatures early in the long tenuous late-phase loop, the bump is absent in the observed AIA 335~{\AA} light curve. Recently, \citet{Zhu18} studied a C-class two-ribbon solar flare using the EBTEL model, in which the modeled peaks in AIA 131 and 94~{\AA} are also well above the observed peaks (Figure~3 in their paper).

On the other hand, the synthetic loop intensities in the cooler AIA 171~{\AA} passband are systematically lower than the observed intensities (Figure~7(e)). It implies an over-estimation of the mass draining from the corona during the radiative cooling phase. The reason may lie in the geometry of the late-phase loop. We note that the best-fit value of parameter $L$ for the southern late-phase loop is 82~Mm in the EBTEL modeling, which is significantly higher than the value of $\sim$55~Mm inferred from the height of the loop apex based on the semi-circular loop assumption. Such a large discrepancy in length is unlikely to be simply ascribed to the uncertainty in determining the loop height, while it is more likely due to an oblate shape of the late-phase loop rather than the semi-circular geometry as commonly assumed. With the same loop length, a semi-circular loop can reach a higher altitude than an oblate loop. Therefore, in a semi-circular loop, the coronal emission is more significantly depressed by the gravitational stratification, which accordingly leads to a higher $c_1$ value. Since the EBTEL model assumes a semi-circular loop geometry for simplicity, it will overestimate the $c_1$ values when dealing with an actually oblate-shaped loop,   and hence inappropriately elevate the mass draining rate during the radiative cooling phase in order to compensate for the over-estimated TR radiative loss \citep{Cargill12}. As a result, the modeled loop intensities during this stage will be under-estimated, as is seen in AIA 171~{\AA} for the late-phase loop modeled here. At higher temperatures, for example, those corresponding to the AIA 335~{\AA} passband, the relative mass draining rate (quantified as $d\ln n/d\ln T$) is not so fast that the relevant loop emission would be much less impacted. In another EBTEL modeling of an M-class solar flare by \citet{LiY12}, the modeled loop intensities in AIA 211 and 171~{\AA} are also systematically lower than the observed intensities (Figure~4 in their paper). We note that \citet{LiY12} fixed the parameter $c_1$  for each individual loop, whose value would be too high for the radiative cooling phase.

\section{Discussion}
Based on the observations and the NLFFF extrapolation results, we propose a two-stage magnetic reconnection scenario to explain the observed flare evolution. The flare is preceded by a brightening first occurring around or within the EW flux rope (manifested by the EW filament), which then propagates eastward and triggers an instability of the nearby NS flux rope (manifested by the NS filament). The details of the triggering process are beyond the scope of this work. Here we conjecture that the helical kink instability may be responsible for the fast ascent of the initially low-lying NS flux rope. As the flux rope rises up, the first-stage magnetic reconnection sets in. Inside the flux rope, possible reconnection due to an internal kink instability \citep{Galsgaard97} can heat the filament body, and more importantly, accelerate high-energy particles that produce the two conjugate HXR sources at the footpoints of the flux rope, as proposed in \citet{Guo12}. Meanwhile, the flux rope pushes the overlying field lines toward the null point, and makes the fan-spine QSL structure thinner, causing an enhancement of the current inside it. Reconnections both around the null point and in the QSL can produce the observed multiple flare ribbons and long late-phase loops connecting the eastern remote ribbon R1 and circular ribbon R2. Considering the elongation of the remote ribbon R1 and the appearance of the late-phase loops, the QSL reconnection seems more energetic than the null-point reconnection. Moreover, since all flare ribbons brighten up very impulsively, we suggest that the QSL reconnection could belong to a super-Alfv{\'e}nic slip-running reconnection \citep{Aulanier06}, during which the inferred elongation motion of ribbon brightening has been complete within the time cadence of AIA\@.

As the fast ascending flux rope slows down, the second-stage magnetic reconnection starts to be initiated. Reconnection mainly takes place in the current sheet formed between the two legs of the overlying field lines stretched by the flux rope. The newly formed ribbon segment R2c fills the gap in the circular ribbon R2 formed in the first stage. Short flaring loops connect R2c with the compact ribbon R3. Such a reconnection is predicted by the standard 2D CSHKP flare model, and the observed separation speed of the flare ribbon ($\sim$30~km~s$^{-1}$ after correction for the projection effect) is also typical for a two-ribbon flare. Therefore, we refer to the observed flaring loops as  ``main flaring loops". Meanwhile, there also exists a less energetic QSL reconnection, which is responsible for the elongation motion of small brightening along the remote ribbon R1. The revealed motion speed of 180~km~s$^{-1}$ implies that the QSL reconnection during this stage may correspond to a sub-Alfv{\'e}nic slipping reconnection \citep{Aulanier06}.

Based on the two-stage reconnection scenario, the long late-phase loops are mainly heated during the first-stage QSL reconnection, while the second-stage flare reconnection, occurring $\sim$5 minutes later, is responsible for the heating of short main flaring loops. This process cannot be accounted for by the standard CSHKP flare model, in which shorter flare loops are always heated earlier than longer loops. Nevertheless, it is a natural consequence of a multipolar magnetic topology, as is the case in this flare. We note that our scenario is basically in agreement with the picture proposed by \citet{SunX13} based on a toy model (Figure~1 in their paper) except that in their opinion the late-phase emission mainly comes from the outer spine lines and is caused by reconnection at the fan-spine null point.

The close correlation between the rise of the flux rope and the brightening of the flare ribbons indicates that the flux rope should be the main driving agent for the reconnections in both stages. During the first stage, the fast ascending flux rope makes the above QSL reconnection more enhanced than usual. Nevertheless, this QSL reconnection does not efficiently remove the strong magnetic constraints above the flux rope, which therefore slows down and finally fails to erupt. The field lines involved in the second-stage reconnection have not yet been sufficiently stretched when the flux rope stops rising, causing a significant depression of the reconnection rate in this stage. Therefore, it is likely that more energy is injected to the late-phase loops than the main flaring loops, leading to an extremely large EUV late phase in this non-eruptive flare.

We also quantify the heating history in a typical late-phase loop. The chromospheric AIA 1600~{\AA} light curve observed at the loop footpoint further confirms that the late-phase loop is heated by an impulsive energy release during the first-stage reconnection. By using the EBTEL model, we synthesize the loop intensities and compare them with observations. Fitting the observed intensities with the synthetic ones yields best-fit values for the peak heating rate $H_0=$1.1~erg~cm$^{-3}$~s$^{-1}$ and loop half-length $L=$82~Mm.  Due to a contamination from the TR and chromosphere, we cannot reliably extract the loop intensities for the much shorter main flaring loops. Furthermore, we cannot reliably characterize the contribution of nonthermal electron beam heating, which is believed to be marginal in the heating of late-phase loops but becomes important during main flare heating. These factors prevent us from applying the same EBTEL modeling to the main flaring loops and making a direct comparison of the heating rate between the two sets of loops. Nevertheless, when comparing our results with those revealed in other flares of the similar \emph{GOES} classes \citep[e.g.,][]{LiY12,Zhu18}, it is found that the peak heating rate of the late-phase loops in this flare is comparable to or even greater than the heating rate of the main flaring loops in those flares. Considering the larger length of the late-phase loop, it is suggested that the late-phase loop undergoes an intense heating with a peak energy deposition rate ($2H_0L$) as high as 1.8$\times$10$^{10}$~erg~cm$^{-2}$~s$^{-1}$ and total energy input of 7.1$\times$10$^{28}$~erg.

For a late-phase loop with a half-length of 82~Mm, the loop emission in AIA 335~{\AA} peaks almost one hour after the impulsive heating. Since there is no obvious additional heating on the late-phase loop, the time delay of the warm coronal emission peak is mainly due to the long cooling process of the late-phase loop, as proposed in \citet{LiuK13}. By comparison, the main flaring loops are significantly shorter (2$L$$\sim$10--25~Mm), and hence evolve more quickly. In a numerical experiment using the EBTEL model, \citet{Dai18} found that with a strong initial loop heating, the hot and warm coronal emissions peak in the conductive cooling and radiative cooling stages, respectively. For the hot coronal emissions, the main flare peak and the late-phase peak temporally overlap because the cooling rate in all loops is very fast during the conductive cooling phase. However, the small difference in cooling rate between the loops of different lengths can still produce an observable bump following the main peak, which results in a dual-decay behavior in the corresponding light curves, as seen in this flare (Figures~1(a) and (c)). We note that this dual-decay in \emph{GOES} SXRs has been used as a proxy for EUV late-phase flares prior to the \emph{SDO} era \citep{Woods14}. When the loops continue to cool down, the effect of the difference in loop length on the loop cooling time becomes more and more prominent. At warm coronal temperatures, the late-phase peak can be well separated and thus easily distinguished from the main flare peak. Since the warm coronal emission peaks in the radiative cooling stage, the emission after the peak can be notably affected by the mass draining, hence showing a fast decay compared to the relatively gradual rise. This pattern is consistent with the shape of the light curves of AIA 335~{\AA} in this flare, both for the whole AR (Figure~1(e)) and for the individual late-phase loop (Figure~7(d)).

In the EBTEL modeling of the late-phase loop, the synthetic peak intensities in the hotter AIA passbands are significantly higher than those observed. We attribute this discrepancy to a non-equilibrium ionization effect \citep{Reale08,Bradshaw11} in the late-phase loop. To further verify this hypothesis, we perform a new set of fitting in which the observational data from both AIA 335 and 94~{\AA} are fitted, as done in \citet{LiY12}. In this approach, no matter how we change the starting guesses for the parameters, the synthetic peak in AIA 94~{\AA} based on the ``best-fit" parameters is still systematically above the observed peak, although the discrepancy between them is somewhat reconciled, and more importantly, the synthetic peak in AIA 335~{\AA} is now systematically lower than the observed peak (not shown here). Since the light curve in neither AIA 335~{\AA} nor AIA 94~{\AA} can be reasonably modeled, the even worse fitting results indicate that the late-phase emissions observed in AIA 94~{\AA} (and also in AIA 131~{\AA}) are indeed affected by the effect of non-equilibrium ionization. This also explains why an EUV late phase as large as that in AIA 335~{\AA} is not observed in a hotter passband, e.g., AIA 94~{\AA}, even though the late-phase peak is clearly separated from the main flare peak (Figure~1(d)).

In the cooler AIA passbands, the higher mass draining rate during the radiative cooling stage can also cause the late-phase peak to be less prominent  (see Figure~1(f)). On the other hand, we find that the synthetic intensities of the late-phase loop in AIA 171~{\AA} are systematically lower than the observed intensities. We propose that the discrepancy may be caused by a semi-circular assumption of the loop geometry in EBTEL, which is not appropriate for the real geometry of the late-phase loop modeled here. 

\section{Summary}
We have analyzed and modeled an M1.2 non-eruptive solar flare on 2011 September 9 that exhibited an extremely large EUV late phase in the warm coronal emissions. Based on an NLFFF extrapolation, we proposed a two-stage magnetic reconnection scenario to explain the evolution of the flare, in which different reconnections are involved and take place in different places. Using the EBTEL model, we modeled the EUV emissions from a late-phase loop. The modeling reveals a high heating rate for the late-phase loop. Our main conclusion is that the extremely large late phase is powered by an intense heating even earlier than the main flare heating, and the delayed occurrence of the late-phase peak is mainly due to the long cooling process of the long late-phase loops. 

The production of the EUV late phase in this flare is different from that in the 2011 September 6 X2.1 eruptive flare not only in timing of the heating process but also in the heating rate. In addition to these two flares, AR NOAA 11283 has produced a series of EUV late-phase flares, which exhibit different evolution patterns in the late-phase emissions. A comparative study of the production of the late phase in these flares will be presented in a following work.
 
\acknowledgements{We are very grateful to the anonymous referee whose valuable comments and suggestions led to a significant improvement of the manuscript. This work was supported by National Natural Science Foundation of China under grants 11533005 and 11733003, and 973 Project of China under grant 2014CB744203. D.Y. is also sponsored by the Open Research Project of National Center for Space Weather, China Meteorological Administration. \emph{SDO} is a mission of NASA's Living With a Star (LWS) Program.}

%\bibliographystyle{aasjournal}
%\bibliography{ms}

\end{document}